\documentclass[aps,preprint]{revtex4}
\def\BibTeX{{\rm B\kern-.05em{\sc i\kern-.025em b}\kern-.T
    08em\kern-.1667em\lower.7ex\hbox{E}\kern-.125emX}}
\usepackage{graphicx}
\usepackage{amsmath}  
\usepackage[utf8]{inputenc}
\usepackage[english]{babel}
\usepackage{amsfonts}
\usepackage{amssymb}

\begin{document}

\title{Magnetic illusion: transforming a magnetic object into another object by negative permeability}

\author{Rosa Mach-Batlle, Albert Parra, Sergi Laut, Nuria Del-Valle, Carles Navau, and Alvaro Sanchez$^{*}$}

\affiliation{Departament de F\'{\i}sica, Universitat Aut\`onoma de Barcelona, 08193 Bellaterra,
Barcelona, Catalonia, Spain}

\begin{abstract}

We theoretically predict and experimentally verify the illusion of transforming the magnetic signature of a 3D object into that of another arbitrary object. This is done by employing negative-permeability materials, which we demonstrate that can be emulated by tailored sets of currents. The experimental transformation of the magnetic response of a ferromagnet into that of its antagonistic material, a superconductor, is presented to confirm the theory. The emulation of negative-permeability materials by currents provides a new pathway for designing devices for controlling magnetic fields in unprecedented ways.

\end{abstract}

\maketitle

\section{Introduction}

\begin{figure}[bt]
	\centering
	\includegraphics[width=0.9\textwidth]{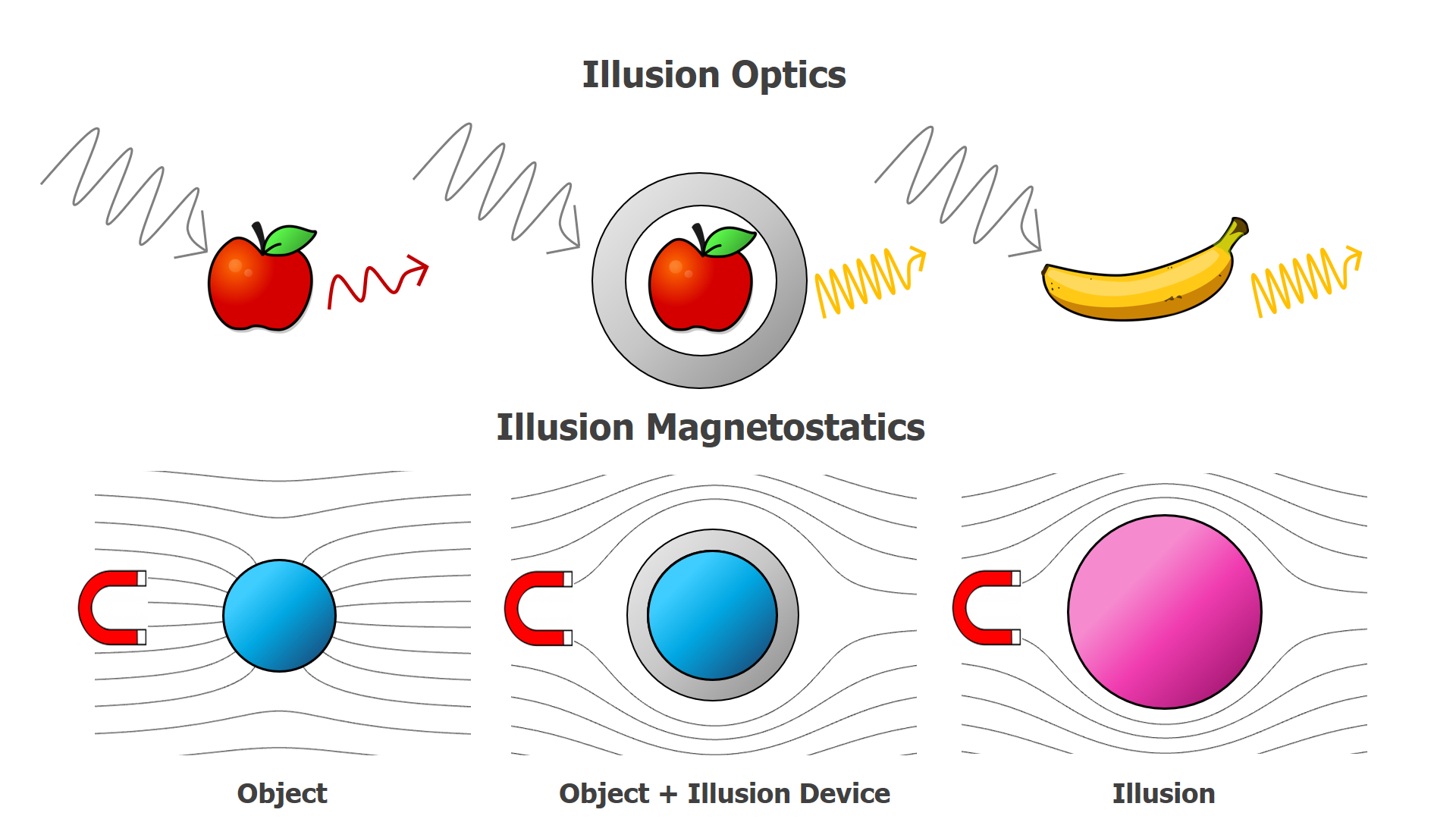}
			\caption{Sketch of the illusions for light (upper row) and magnetic fields (lower row). In the upper row, the image of an object (an apple) illuminated by light is transformed into that of another object (a banana) by surrounding the former with an illusion device (in grey color), as in \cite{illusion}. Analogously, in the lower row, the magnetic response of an object (a ferromagnetic material, attracting fields lines) is transformed by using a magnetic illusion device (in grey color) into that of another magnetic object (a perfectly diamagnetic one, repelling field lines).  } 
	\label{illusion scheme}
\end{figure}

The physical appearance of objects is essential for our apprehension of reality and also for the technologies relying on the identification of objects based on their signature.
Illusion, the transformation of the appearance of an object into that of another one, is an active research field \cite{illusion,jiang,yang,xiang,bilotti,eleftheriades}. The illusion of transforming the response of an object to impinging electromagnetic waves (Fig. \ref{illusion scheme}) has been theoretically studied using the tools of transformation optics \cite{illusion} and experimentally demonstrated in reduced schemes
 \cite{illusion_circuit,illusion_radar,illusiondc,eleftheriades2}. Such optical illusions typically require media with negative values of the refraction index. These media were first introduced in  
seminal works by Veselago \cite{veselago} and Pendry {\it et al.} \cite{pendry} and gave rise to a host of new ways of controlling and manipulating light propagation. The theoretical predictions were experimentally demonstrated with the help of metamaterials, artificial materials with unusual electromagnetic properties that are not found in naturally occurring materials \cite{smith,tjc_book,zheludev,shalaev}.

Magnetic metamaterials \cite{wood,Magnus2008,antimagnet} have recently been developed, enabling the realization of novel devices for controlling magnetic fields \cite{antimagnet,narayana,gomory,genenko,concentrator,SunHe,AnlageSUST2014,hose,wormhole,3dcloak}. They include cloaks that make objects magnetically undetectable \cite{antimagnet,narayana,gomory,genenko,3dcloak} and shells that concentrate magnetic energy in a given spatial region \cite{concentrator,SunHe}.  However, illusion has not been realized for static magnetic fields until now. The main obstacle for this realization has been the absence of passive materials with negative static permeability \cite{dolgov}.If such negative-$\mu$ materials were available, they would  allow the realization of novel tools for controlling magnetic fields, including the illusion of transforming the magnetic signature of an object into other ones. Such magnetic illusions could enable the transformation of an original object into: i) a different magnetic object (e. g. disguising a ferromagnet as a superconductor), ii) the same material with different dimensions (enlarging or shrinking the original object \cite{superscatterer,transformation}), iii) air (making the original object magnetically invisible \cite{antimagnet,narayana,gomory,genenko,3dcloak}).

In this work we theoretically and experimentally demonstrate illusion for magnetic fields by transforming the magnetic signature of a 3D magnetic object into that of another object of different magnetic nature. The magnetic illusion is achieved by assuming a hypothetical negative-$\mu$ material; this material can be effectively emulated in practice by a suitably tailored set of currents \cite{negativemu}. We experimentally confirm the theoretical ideas by  demonstrating the transformation of extreme antagonistic materials: the response of an ideal ferromagnetic ($\mu\to\infty$) sphere is transformed into the response of a perfect diamagnetic ($\mu\to 0$) one. A sketch of this magnetic illusion is shown in Fig. \ref{illusion scheme}, together with its optical illusion counterpart \cite{illusion}. We also discuss how other illusions, such as magnification and cloak can similarly be achieved assuming negative-$\mu$ materials.

\section{Magnetic illusion}

Consider a sphere of radius $R_1$ and relative magnetic permeability $\mu_1$  in a uniform magnetic induction $B_{\rm a}$ applied along the $z$-direction. We aim at transforming its magnetic response into that of another sphere of radius $R$ and permeability $\mu$. To this purpose, we surround the original sphere with a spherical shell of inner and outer radii $R_1$ and $R_2$, respectively, and permeability $\mu_2$, which acts as the illusion device. We consider that all involved materials are linear, homogeneous, and isotropic; interestingly, illusion solutions are achieved even within these assumptions.  Analytic expressions for the magnetic field in all the space are obtained by solving magnetostatic Maxwell equations (see Supplemental Material \cite{supp} for the derivations). 

To achieve the illusion, the magnetic signature of  the original object surrounded by the illusion device has to be indistinguishable from that of the target object. 
Since we are considering spherical objects, the magnetic field they create as a response to a uniform applied field is equivalent to that of a point dipole. By equating the dipolar magnetic moment of the original sphere plus the illusion device to that of the target sphere, we obtain the general condition 
\begin{equation}
\frac{(\mu_1 - \mu_2) (1+2\mu_2) + (-1+\mu_2)(\mu_1+2\mu_2) \left(R_2/R_1\right)^3 }{2(\mu_1-\mu_2)(-1+\mu_2)+ (2+\mu_2) (\mu_1+2\mu_2) \left(R_2/R_1 \right)^3} R_2^3  =\frac{\mu  - 1}{\mu + 2}R^3.
\label{illusioneq}
\end{equation}
This relation implies that given an original object (with $R_1$ and $\mu_1$) and a target one (with $R$ and $\mu$), one has the freedom of fixing $R_2$ or $\mu_2$ for the illusion device. 
In the case $R=R_2$, a similar relation as Eq. (\ref{illusioneq}) can also be obtained from a generalized effective-medium theory for electromagnetic waves in the limit of long wavelengths \cite{slovick}.

\begin{figure*}[htb]
	\centering
	\includegraphics[width=0.9\textwidth]{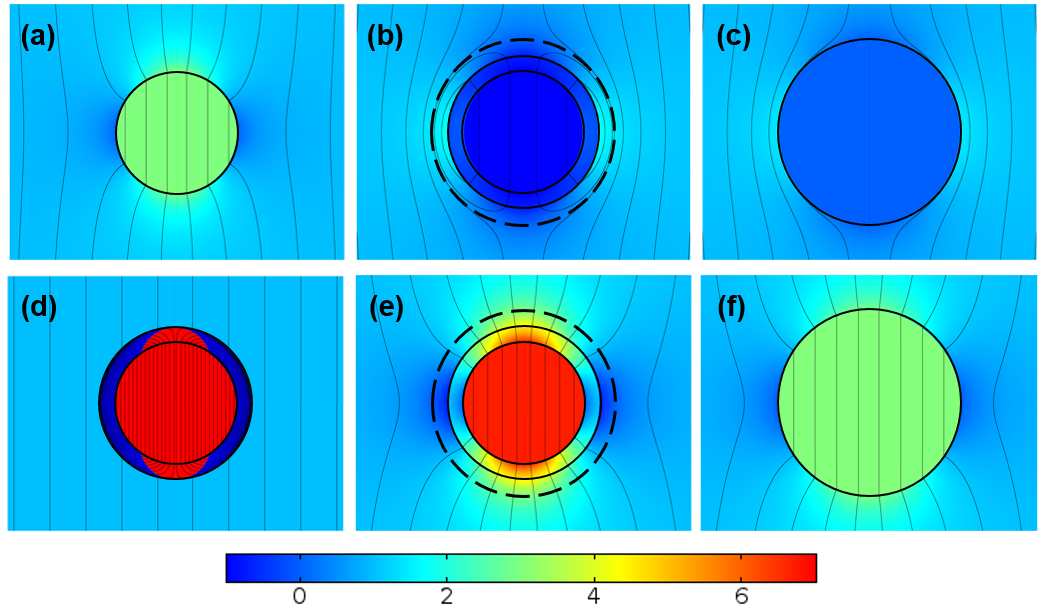}
			\caption{{Magnetic illusions: transformation, cloaking and magnification}. Magnetic induction field lines and its vertical component, $B_z$, resulting from a vertically applied uniform magnetic induction $B_{\rm a}$, normalized to $B_{\rm a}$ (in color scale), for the cases of {\bf (a)} an ideal ferromagnetic  sphere ($\mu_1=1000$, radius $R_1$), {\bf (b)} the sphere in {\bf (a)} surrounded by a spherical-shell illusion device ($\mu_2 = -0.0942$, extending from $R_1$ to $R_2=1.25R_1$),  {\bf (c)} an ideal diamagnetic sphere ($\mu = 1/1000$, radius $R=1.5R_1$), {\bf (d)} the sphere in {\bf (a)} surrounded by a spherical-shell cloaking device ($\mu_2 = -2071$, extending from $R_1$ to $R_2=1.25R_1$), {\bf (e)} the sphere in {\bf (a)} surrounded by a spherical-shell magnifying device ($\mu_2 = -1.47$, extending from $R_1$ to $R_2=1.25R_1$) and {\bf (f)} an ideal ferromagnetic  sphere ($\mu_1=1000$, radius $1.5R_1$). The magnetic signatures in {\bf (b)} and {\bf (e)} are indistinguishable from those in {\bf (c)} and {\bf (f)}, respectively, in the region of illusion, the inner border of which is the dashed line in {\bf (b)} and {\bf (e)}.} 
	\label{fig_theory}
\end{figure*}

Several interesting cases of magnetic illusion directly result from Eq. \eqref{illusioneq}; the analytic field distributions for some examples are shown in Fig. \ref{fig_theory}. A first one is transforming the object into another of different magnetic nature, $\mu \neq \mu_1$. An original ferromagnetic sphere, Fig. \ref{fig_theory}(a), surrounded by the illusion shell, Fig. \ref{fig_theory}(b), has the same magnetic response as a superconducting sphere, Fig. \ref{fig_theory}(c). A second example is cloaking, that is, transforming the magnetic signature of the original object into that of empty space, $\mu=1$, rendering the object magnetically undetectable. In Fig. \ref{fig_theory}(d) the original ferromagnetic sphere is made magnetically invisible by a cloaking shell. A third example is the magnification, $R > R_1$, or shrinking, $R < R_1$, of the original object, $\mu = \mu_1$. The ferromagnetic sphere surrounded by a magnifying shell, Fig. \ref{fig_theory}(e), appears as an enlarged object, Fig. \ref{fig_theory}(f).
Similar effects have been demonstrated for microwaves \cite{eleftheriades2,shinkring} and dc electric fields \cite{switchable}, and theoretically predicted for electromagnetic waves \cite{bilotti} and magnets \cite{transformation}.

\section{Negative magnetostatic permeabilty}
On closer inspection, Eq. (\ref{illusioneq}) reveals why illusion has never been realized in magnetostatics until now; the illusion device requires in general negative values of the permeability $\mu_2$. Whereas effective negative permeabilities have been obtained for electromagnetic waves at several frequency ranges, typically by ressonant elements \cite{smith,alunegative}, negative permeability has not been realized by passive materials in magnetostatics \cite{dolgov}. However, we circumvent this limitation here by using the property that in magnetostatics the magnetic response of a material can be emulated by substituting the material with its magnetization currents, $\bold{J_M} = \nabla \times \bold{M}$ in the bulk and $\bold{K_M} = \bold{M} \times \bold{n}$ on the surfaces, where {\bf M} is the magnetization and $\mathbf{n}$ is the vector perpendicular to the surface  \cite{negativemu}. Therefore, by externally supplying the adequate set of currents one can exactly emulate the behavior of a material with negative $\mu$ without actually having it.

\section{Transforming a ferromagnet into a superconductor}

To demonstrate the potential of negative-$\mu$ materials, we experimentally realize the illusion of transforming the magnetic signature of a ferromagnetic sphere into that of its antagonistic material, a perfect diamagnetic (e.g. superconducting) one. 

We choose the radii relations $R_2 = 1.25 R_1$ and $R = 1.5 R_1$. In general, when fixing the radii, Eq. (\ref{illusioneq}) has two solutions for $\mu_2$. Only when the original sphere has extreme values ($\mu_1\to \infty$ or $\mu_1\to 0$) a single solution exists. In our case, Eq. (\ref{illusioneq}) leads to $\mu_2=-0.0942$, which means that the desired illusion device has negative permeability. This illusion is displayed in Fig. \ref{fig_theory}(a)-(c).

We next find the currents that emulate the required illusion device. Because $\mu_1$ and $\mu_2$ are homogeneous and isotropic, the volume magnetization currents $\mathbf{J_{M}}$ are zero. The surface magnetization current densities $\mathbf{K_{M_1}}$ and $\mathbf{K_{M_2}}$, flowing at the surfaces $r=R_1$ and $r=R_2$, respectively, would give the same response as the system constituted by the inner ferromagnetic sphere plus the illusion device. Since our setup includes an actual ferromagnetic sphere, the currents $\mathbf{K_{M_1}}$ flow on its surface. Therefore, to emulate the illusion device there is no need to supply $\mathbf{K_{M_1}}$ 
but only the current density
\begin{equation}
\mathbf{K_{M_2}}(\theta) =-{3\over{2}}\left( \frac{2R_1^3+R^3}{2 R_1^3 + R_2^3}\right){B_a\over{\mu_0}} {\rm sin}\theta  \mathbf{e_\varphi}.  
\label{KM2}
\end{equation}
These currents create the desired illusion taking into consideration that the ferromagnet responds not only to the applied field but also to the field created by the currents themselves.

\begin{figure*}[htb]
	\centering
	\includegraphics[width=0.9\textwidth]{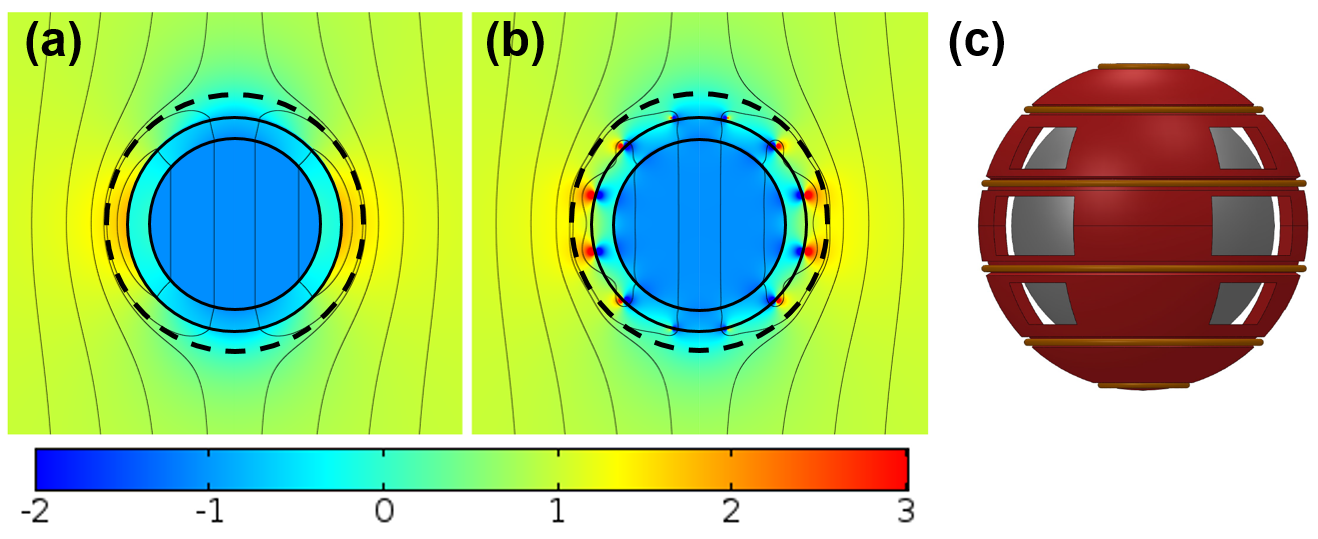}
			\caption{Magnetic induction field lines and its vertical component, $B_z$, resulting from a vertically applied uniform magnetic induction $B_{\rm a}$, normalized to $B_{\rm a}$ (in color scale), for the cases of {\bf (a)} an ideal ferromagnetic  sphere ($\mu_1=1000$, radius $R_1$) surrounded by a spherical-shell illusion device ($\mu_2 = -0.0942$, extending from $R_1$ to $R_2=1.25R_1$), and {\bf (b)} same as {\bf (a)} with a discretized current distribution of the illusion device. {\bf (c)} Sketch of the ferromagnetic sphere surrounded by the current loops composing the illusion device used in the experiments. } 
	\label{illusion_disc}
\end{figure*}

In the experimental realization, the surface current distribution of Eq. \eqref{KM2} is substituted by a discrete set of 6 loops of current [see Fig. \ref{illusion_disc}(c)]. Finite-element simulations demonstrate that the response of this discretized system [Fig. \ref{illusion_disc}(b)] is very similar to that of the ideal system [Fig. \ref{illusion_disc}(a)].

\section{Experimental realization}

\begin{figure}[thb]
	\centering
		\includegraphics[width=0.6\textwidth]{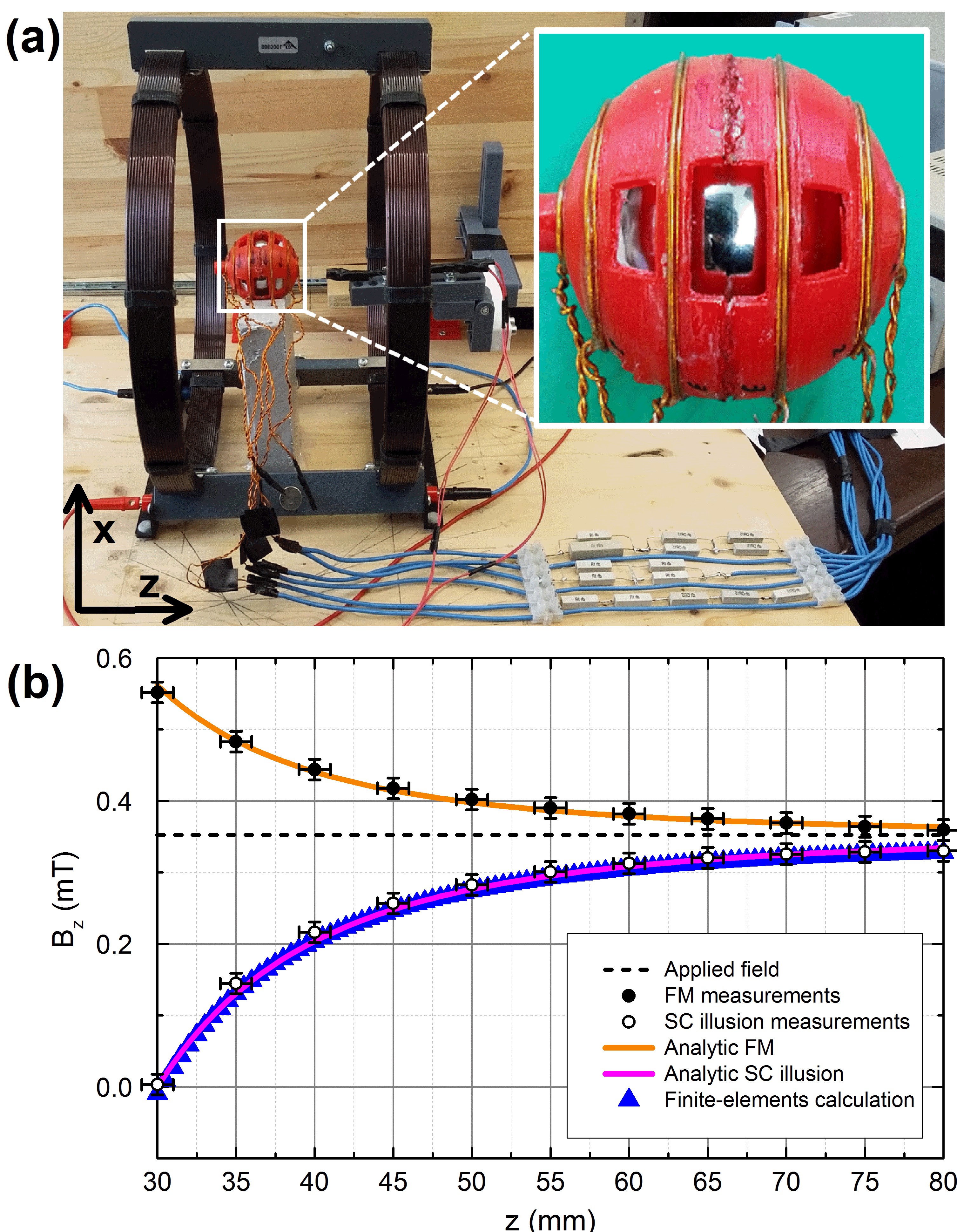}
	\caption{{\bf (a)} Experimental setup with a picture of the discretized illusion device with the current loops on a non-magnetic 
3D-printed holder surrounding the ferromagnetic sphere. {\bf (b)} Measured magnetic induction $B_z$ along the $z$-axis. Open and solid circle symbols correspond to  the complete illusion setup (SC) and to the ferromagnetic-only (FM) cases, respectively. Solid lines are the analytic results for the target superconducting sphere (magenta) and the original ferromagnetic one (orange). Finite-elements calculations for the discretized device are shown with triangular symbols.
Dashed line indicates the applied $B_{a,z}$ value.	
} 
	\label{Experiments}
\end{figure}

To realize the experiments, we surround a solid steel sphere of radius $R_1 = 20$mm with a non-magnetic spherical 3D-printed former of radius $R_2 = 25$mm. The radius of the target diamagnetic object is $R =30$mm. 
The 6 current loops of the discretized illusion device are placed in grooves on the former, as shown in Fig. \ref{illusion_disc}(c) and Fig. \ref{Experiments}(a). The loops are connected to the same power supply, in parallel to each other. The required current at each loop is fed by connecting resistors in series.
This setup is placed at the central region of a pair of Helmholtz coils [see Fig. \ref{Experiments}(a)] that provides a constant applied magnetic induction along the $z$-direction, $B_{a,z}=0.353$mT, taking into account the contribution of the Earth magnetic field.
The $z$-component of {\bf B} is measured with a Hall probe, starting from $r=R$ along both the $z$-axis and the $x$-axis.   Measurements are performed for both the original ferromagnetic sphere and the complete illusion scheme, as shown in Fig. \ref{Experiments}(b). 
Very good agreement between experimental and theoretical results is obtained for both the complete setup and the ferromagnet solely. Results demonstrate that we have successfully transformed the magnetic response of a ferromagnet into that of a superconductor, even using a discretized illusion device.
In this way, an effective superconducting response has been obtained, without actually employing superconducting materials, in the presence of other magnetic objects. Further experimental confirmation for another illusion case with different sphere dimensions is shown in the Supplemental Material \cite{supp}.

\section{Discussion}
The general framework we have constructed based on emulating a hypothetical negative-$\mu$ material with currents allows the design of novel devices for controlling magnetic fields. One interesting application is the illusion of magnifying a magnetic object, as shown in Fig. \ref{fig_theory}(e)-(f). Using negative values of permeability in the illusion device, the apparent radius of the object can be made much larger than the external radius of the illusion device, something unachievable with positive permeability values. The consideration of negative permeability introduces a new solution to cloak the response of a magnetic object, extending the known solutions with positive permeability \cite{gomory,wormhole,3dcloak}. An example of the new negative-permeability cloaking solutions is shown in Fig. \ref{fig_theory}(d).
Only when the permeability of the original object, $\mu_1$, has an extreme value (either $\mu_1\to 0$ or $\mu_1\to \infty$)  there is a single cloaking solution, the positive-permeability one; this kind of solutions has been experimentally demonstrated for a superconductor-ferromagnetic bilayer magnetic cloak in \cite{gomory,wormhole,3dcloak} and in other areas involving static fields, like diffusive media \cite{Schittny} or thermal effects \cite{xu,han}.

Since negative-permeability media can be emulated by supplying energy in the form of currents - which are simply proportional to the applied field [Eq. (\ref{KM2})], these materials can be understood as active \cite{alu_review,green,active_thermal,
active_tiejuncui,eleftheriades,eleftheriades2}. 
This fact makes our illusion device fully controllable - e. g. it can be easily switched on and off at will. 
By a dynamic control of the currents one could even transform the magnetic signature of an initial object successively into that of different ones, or, for example, turn a cloak on and off. In the present realization, our illusion is predetermined \cite{miller} because the currents are specific for the applied field value. One could make a magnetostatic illusion without prior knowledge of the applied field by using a set of sensors and a feedback loop as in \cite{miller, negativemu}.

It is interesting to compare our magnetostatic illusions to those studied for the full electromagnetic wave case. 
One strategy for illusion for waves typically involves two different media, a complementary media and a restoring one, both involving complicated distributions of permittivity and permeability, often obtained by transformation optics \cite{illusion,TO}. Here we have shown how magnetostatic illusions can be realized in a much simpler way, surrounding the original object by a single homogenous and isotropic medium whose response can be emulated by currents. Our strategy is reminiscent to the active electromagnetic cloaks and illusions achieved using the equivalence principle \cite{eleftheriades,eleftheriades2}. In these works, superimposing magnetic and electric surface current densities at the boundary of an object enables the control of scattered fields. However, if this procedure was applied to our magnetostatic case, the interaction of the material with the magnetic field created by the electric currents would spoil the desired effect - e.g. illusion or cloak. In our illusion device, thanks to the introduction of negative permeability, the current distribution is obtained taking into account its interaction with the material.

The results confirm the static limit as a very promising case for applying the ideas of transforming or disguising materials into different ones; a recent example is the transformation of the sign of the Hall effect in semiconductors by metamaterials \cite{kern}. Another advantage of the static case is that, since there is no characteristic wavelength involved, the illusions can be made at any scale. Our results can also be valid in the quasistatic limit, for low frequency electromagnetic waves such that the wavelength is far larger than the device dimensions. In this limit the negative permeability illusion device could be emulated by ac current loops.

In this work we have limited our study to the illusion of spherical objects. For long cylinders in transverse fields, which also generate a dipolar response, analytic expressions analogous to Eq. (\ref{illusioneq}) can be found. Our method for illusion could also be applied to geometries for which the original and/or the target magnetic object generate a multipolar response (e. g. a cube). In these cases each term of the multipolar response of the original object would have to be transformed into that of the target object. This could require in general an inhomogeneous permeability in the illusion device, which could always be converted into equivalent sets of magnetization currents.

\section{Conclusion}

In summary, we have shown that negative magnetostatic permeability is a general tool for designing novel magnetic field configurations.  We have demonstrated that the effective properties of negative-$\mu$ materials can be achieved by suitably tailored sets of currents. As a relevant example, we have experimentally realized the illusion of transforming the magnetic signature of a ferromagnet into that of a superconductor.  
Other phenomena such as magnification and cloaking have been shown to be realizable using the same ideas. 
Because magnetic signatures are essential in many applications, from the reading and writing of magnetic bits in computer memories or credit cards to the magnetic traces left by hidden objects in security detectors, the transformation of a given magnetic signature into a different one may be relevant in current and future technologies.

\section*{Acknowledgements}
We thank Jordi Prat-Camps for discussions and help in the experiments. We also thank European Union Horizon 2020 Project FET-OPEN MaQSens (grant agreement 736943), and
projects MAT2016-79426-P (Agencia Estatal de Investigación / Fondo Europeo de Desarrollo Regional) and 2017-SGR-105
for financial support.  A. S. acknowledges a grant from ICREA Academia, funded by the Generalitat de Catalunya.

\end{document}